\documentclass{ieeeaccess}
\usepackage{cite}
\usepackage{amsmath,amssymb,amsfonts}
\usepackage{algorithmic}
\usepackage{graphicx}
\usepackage{textcomp}
\usepackage{array}
\usepackage{epstopdf}
\usepackage{ulem}
\def\BibTeX{{\rm B\kern-.05em{\sc i\kern-.025em b}\kern-.08em
    T\kern-.1667em\lower.7ex\hbox{E}\kern-.125emX}}
\begin{document}
\history{Date of publication xxxx 00, 0000, date of current version xxxx 00, 0000.}
\doi{10.1109/ACCESS.2019.DOI}

\title{Information flow networks of Chinese stock market sectors}

\author{\uppercase{Peng Yue}\authorrefmark{1,2}, 
\uppercase{Qing Cai\authorrefmark{1,2}, Wanfeng Yan\authorrefmark{2,3}, and Wei-Xing Zhou}\authorrefmark{1,2,4}}
\address[1]{School of Business, East China University of Science and Technology, Shanghai 200237, China}
\address[2]{Research Center for Econophysics, East China University of Science and Technology, Shanghai 200237, China}
\address[3]{Zhicang Technologies, Beijing 100016, China}
\address[4]{Department of Mathematics, East China University of Science and Technology, Shanghai 200237, China}

\tfootnote{This work was partly supported by the National Natural Science Foundation of China (U1811462), the Shanghai Philosophy and Social Science Fund Project (2017BJB006), the Program of Shanghai Young Top-notch Talent (2018), and the Fundamental Research Funds for the Central Universities.}

\markboth
{P. Yue \headeretal: Information flow networks of Chinese stock market sectors}
{P. Yue \headeretal: Information flow networks of Chinese stock market sectors}

\corresp{Corresponding authors: Wanfeng Yan (e-mail: wanfeng.yan@gmail.com) or Wei-Xing Zhou (e-mail: wxzhou@ecust.edu.cn)}

\begin{abstract}
Transfer entropy measures the strength and direction of information flow between different time series. We study the information flow networks of the Chinese stock market and identify important sectors and information flow paths. This paper uses the daily closing price data of the 28 level-1 sectors from Shenyin \& Wanguo Securities ranging from 2000 to 2017 to study the information transmission between different sectors. We construct information flow networks with the sectors as the nodes and the transfer entropy between them as the corresponding edges. Then we adopt the maximum spanning arborescence (MSA) to extracting important information flows and the hierarchical structure of the networks. We find that, during the whole sample period, the \textit{composite} sector is an information source of the whole stock market, while the \textit{non-bank financial} sector is the information sink. We also find that the \textit{non-bank finance}, \textit{bank}, \textit{computer}, \textit{media}, \textit{real estate}, \textit{medical biology} and \textit{non-ferrous metals} sectors appear as high-degree root nodes in the outgoing and incoming information flow MSAs. Especially, the \textit{non-bank finance} and \textit{bank} sectors have significantly high degrees after 2008 in the outgoing information flow networks. We uncover how stock market turmoils affect the structure of the MSAs. Finally, we reveal the specificity of information source and sink sectors and make a conclusion that the root node sector as the information sink of the incoming information flow networks. Overall, our analyses show that the structure of information flow networks changes with time and the market exhibits a sector rotation phenomenon. Our work has important implications for market participants and policy makers in managing market risks and controlling the contagion of risks.
\end{abstract}


\titlepgskip=-15pt

\maketitle

\section{Introduction}
\label{S1:Intro}

Complex systems are usually composed of some interrelated subsystems and understanding the interactions between different subsystems makes a lot of sense. A network representation is found advanced to characterize the complex systems and has been applied far and wide in many scientific fields, such as stock networks \cite{Tse-Liu-Lau-2010-JEF,Battiston-Rodrigues-Zeytinoglu-2007-ACS}, communication networks \cite{Bogachev-Bunde-2009-EPL}, Internet \cite{PastorSatorras-Vazquez-Vespignani-2001-PRL}, World Wide Web \cite{Albert-Jeong-Barabasi-1999-Nature}, and economic networks \cite{Schweitzer-Fagiolo-Sornette-VegaRedondo-Vespignani-White-2009-Science,Mantegna-1999-EPJB,Onnela-Kaski-Kertesz-2004-EPJB,EmmertStreib-Musa-Baltakys-Kanniainen-Tripathi-YliHarja-Joblbauer-Dehmer-2018-JNTF}, to list a few. Financial markets, as a representative type of complex systems, reflect a dynamic interaction of a large number of different elements at different levels, including different traders, stocks and sectors, etc. The financial crises such as those in China since 2000 (June 2001 \cite{Zhou-Sornette-2004a-PA}, December 2007 \cite{Jiang-Zhou-Sornette-Woodard-Bastiaensen-Cauwels-2010-JEBO}, June 2009 \cite{Jiang-Zhou-Sornette-Woodard-Bastiaensen-Cauwels-2010-JEBO} and June 2015 \cite{Sornette-Demos-Zhang-Cauwels-Filimonov-Zhang-2015-JIS}) demonstrate a critical demand for a fundamental and new understanding of the structure and dynamics evolutions of financial market networks. 

There are many techniques have been proven effective to construct networks for financial markets, such as correlation analysis \cite{Bonanno-Lillo-Mantegna-2001-QF,Plerou-Gopikrishnan-Rosenow-Amaral-Guhr-Stanley-2002-PRE,Onnela-Kaski-Kertesz-2004-EPJB,Bonanno-Caldarelli-Lillo-Micciche-Vandewalle-Mantegna-2004-EPJB,Wang-Xie-Chen-2017-JEIC,Wang-Xie-Stanley-2018-CE,Zhang-Zhang-Shen-Zhang-2018-Complexity}, Granger causality \cite{Jordn-Nick-2000-AFE,Yao-Lin-Lin-2016-PA}, mutual information \cite{Dionisio-Menezes-Mendes-2004-PA,Abigail-2013-PA,Pawel-2014-PRE,Baitinger-Papenbrock-2017-JNTF}, tail dependence \cite{Wang-Xie-2016-ESA,Wen-Yang-Zhou-2019-IJFE}, and so on. Compared to undirected networks or directed but unweighted networks, directed and weighted networks can characterize better the interaction between subsystems and the structure of complex systems. Information flow theory is characterized by interaction and has been widely adopted in analyzing economic systems \cite{DHulst-Rodgers-2000-IJTAF,Eguiluz-Zimmermann-2000-PRL}. The main idea of the interaction is the direction and strength of coupling. In this study, we use transfer entropy to identify the information transfers between different industrial sectors in the Chinese stock market and construct an information flow network. Transfer entropy (TE), as a kind of log-likelihood ratio \cite{Barnett-Bossomaier-2012-PRL}, is a measurement method that quantifies information flow based on the probability density function (PDF). It cannot only identifies the direction of the information flow but also quantifies the flows between different subsystems, which has been widely applied \cite{Schreiber-2000-PRL, Ai-2014-Entropy, Hu-Zhao-Ai-2016-Entropy, Yook-Chae-Kim-Kim-2016-PA, Javer-Nicola-Bruno-Sandra-Alex-Yamir-Alessandro-2016-SciAdv,Zhang-Lin-Shang-2017-FNL, Toriumi-Komura-2018-CE, Servadio-Convertino-2018-SciAdv,Zhang-Shang-Xiong-Xia-2018-FNL}. 

There are also many scholars study the impact of financial crisis on the structure of stock markets from the perspective of information flow network. Oh et al. investigated the information flows among different sectors of the Korean stock market \cite{Oh-Oh-Kim-Kwon-2014-JKPS}. They measured the amount of information flow and the degree of information flow asymmetry between industrial sectors around the sub-prime crisis and showed clearly that the insurance sector as the key information source after the crisis. Sandoval Jr. built dynamic networks based on correlation and transfer entropy, unveiling node strength peaks in times of crisis \cite{Sandoval-2017-JNTF}. Gon{\c{c}}alves and Atman used the visibility graph algorithm combined with information theory to construct an estimator of stock market efficiency for the impact caused by the 2008 European crisis in the global markets \cite{Goncalves-Atman-2017-JNTF}. Wang and Hui applied effective transfer entropy to study the information transfer in the Chinese stock market around its crash in 2015. They divided the crash into the tranquil, bull, crash, and post-crash periods and they found the information technology sector is the biggest information source, while the consumer staples sector receives the most information \cite{Wang-Hui-2018-Entropy}. Li et al. used transfer entropy to research the risk contagion in Chinese banking system, and evaluated the stability of Chinese banking system by simulating the risk contagion process \cite{Li-Liang-Zhu-Sun-Wu-2013-Entropy}. To extend these works, we detect the evolution of out-degree and in-degree of the root sector during the crisis period, and then we expose the key sector in our information flow networks between pre-crisis and post-crisis periods.

How to extract important information and understand the properties of the aforementioned information flow networks become more difficult and crucial since these networks are usually complete graphs which means all links between pairs of nodes are present. There are a variety of filtering tools applied by different researchers, such as Planar Maximally Filtered Graph (PMFG) \cite{Tumminello-Aste-DiMatteo-Mantegna-2005-PNAS}, Minimum (or maximum) Spanning Tree (MST) \cite{Micciche-Bonanno-Lillo-Mantegna-2003-PA} and bootstrap reliability estimation \cite{Tumminello-Coronnello-Lillo-Micciche-Mantegna-2007-IJBC}. Among them, MST is an efficient tool to detect the hierarchical structure in financial markets \cite{Mantegna-1999-EPJB}, which has been adopted in many works \cite{Onnela-Chakraborti-Kaski-Kertesz-Kanto-2003-PRE,Heimo-Saramaki-Onnela-Kaski-2007-PA,Heimo-Kaski-Saramaki-2009-PA,Eryigit-Eryigit-2009-PA,Eom-Oh-Jung-Jeong-Kim-2009-PA,Gilmore-Lucey-Boscia-2010-PA,Dias-2012-PA,Kumar-Deo-2012-PRE,Yang-Shen-Xia-2013-MPLB,Wang-Xie-Chen-Chen-2013-Entropy,Colrtti-2016-PA,Cao-Zhang-Li-2017-CSF,Kazemilari-Mardani-Streimikiene-Zavadskas-2017-RE}. Kwon and Yang studied the strength and direction of information flow between 25 stock indices using transfer entropy and minimum spanning tree. They found that the Standard and Poor's 500 Index is located at the information source in the global stock market \cite{Kwon-Yang-2008-EPL}. We note that the directed analog of maximum spanning tree is usually called maximum spanning arborescence (MSA). To the best of our knowledge, there is no work combining these two methods TE and MST (or more specifically MSA) to analyze the evolution of the information flow networks of different industrial sectors in financial markets. Hence, we construct transfer entropy based networks and use MSA to conduct a dynamic analysis to give some new insights into the Chinese stock market.

It has been a ``stylized fact'' that the stock price can be considered as a barometer of the company reflecting the sentiment of participants about the company. Similarly, the stock index price can be deemed as a barometer of a series of companies which belong to the same economic sector. We aim to reveal the interrelationship between different economic sectors of the Chinese stock market and the impact of market crashes on its structure. To this end, in our empirical analysis, we first choose 28 level-1 stock indices issued by Shenyin \& Wanguo Securities to construct transfer entropy based information flow networks, which are directed and weighted. Next, by employing the filtering technique of MSA, we identify the important information flow propagation paths and the key sectors of the incoming and outgoing information flow networks in different years. Finally, we investigate how the two market crashes (occurred in 2008 and 2015) impact the structure of the information flow network. 

Overall, this paper contributes to the literature on information flow networks in several ways. {First, we identify and analyze the paths of information flow transfers among different sectors. In particular, we investigate the role of different sectors in the information flow networks such as root sectors, leaf sectors and the sectors located in the maximal information flow paths, which have important implications in risk management for policy makers and strategy design for market participants. Second, we compare not only the relationship between the degree of root sectors and the status of the stock market but also the relationship between the amount information of MSAs and the status of the stock market. Third, such an analysis has not been conducted on the Chinese stock market, which is becoming more and more important in the global market and attracting increasing interest.}

The remainder of this paper is organized as follows. Section \ref{S1:Method} describes the minimum spanning tree approach, the method for calculating transfer entropy, and how we use it to construct information flow networks. Section \ref{S1:Data} overviews the sector index time series for the Chinese stock market. Section \ref{S1:Results} presents the empirical results about the evolution of MSAs of the information flow networks and how financial crisis affect the shape of the MSAs. Section \ref{S1:Conclude} concludes this work and provides some new insights to the Chinese stock market.

\section{Methods}
\label{S1:Method}

In this section, we describe the main methods adopted in this work. The analyses are implemented using MatLab, with which our university (ECUST) has a license. Certainly, softwares other than MatLab can also do the job.

\subsection{Symbolic transfer entropy}

Transfer entropy is a useful method in information theory. Schreiber is the first to use transfer entropy to measure information transfer and detect asymmetry in the interactions among subsystems \cite{Schreiber-2000-PRL}. To explore the transfer entropy between two time series, there are a variety of approaches in the literature. We use the symbolic transfer entropy introduced by Staniek and Lehnertz \cite{Staniek-Lehnertz-2008-PRL}.

Consider two time series of daily closing prices of two sector indices \{$Y_t$\} and \{$X_t$\}, $t=1,2,\ldots, L$. The symbolic transfer entropy $T^S_{y\rightarrow{x}}$ from the corresponding return time series \{$y_t$\} to \{$x_t$\} can be calculated as follows. 
First, we calculate the logarithmic returns \{${x_t}$\} of the original price time series \{$X_t$\}, which are calculated as follows
\begin{equation}
x_t\equiv{\ln(X_t)-\ln(X_{t-1})}.
\end{equation}
Then the returns are discretized into $q$ non-overlapping windows of identical length $\Delta$. The width of each window is $\Delta_x=[x_{\max}-x_{\min}]/q$ and the $k$th window is $[x_{\min}+(k-1)\Delta_x, x_{\min}+k\Delta_x)$, where $x_{\max}$ and $x_{\min}$ are respectively the maximum and minimum values of the time series $x_t$. Similarly, we repeat the procedure for $y_t$, in which $\Delta_y$ is usually different from $\Delta_x$.
Next, we map the logarithmic return time series to $\hat{x}$ and $\hat{y}$ by $\hat{x}_t=f(x_t)=k_x$ and $\hat{y}_t=f(y_t)=k_y,$ where $k_x,k_y=1,2,\cdots,q,~x_t\in{[x_{\min}+(k_x-1) \Delta_x, x_{\min}+k_x \Delta_x)}$ and $y_t\in{[y_{\min}+(k_y-1) \Delta_y, y_{\min}+k_y \Delta_y)}$.
Finally, we count the numbers of returns in the $q$th window, denoted by $\hat{x}_t^q$ and $\hat{y}_t^q$ respectively, and then calculate the probabilities $p(\hat{x}_t)=\hat{x}_t^q/(L-1)$ and $p(\hat{y}_t)=\hat{y}_t^q/(L-1)$, as well as the joint probabilities $p(\hat{x}_t,\hat{y}_t)$, $p(\hat{x}_t,\hat{x}_{t+1})$ and $p(\hat{x}_{t+1},\hat{x}_t,\hat{y}_t)$. The symbolic transfer entropy from $\{{y_t}\}$ to time series $\{{x_t}\}$ is calculated as
\begin{equation}
T^S_{y\rightarrow{x}}=\sum_{\hat{x}_{t+1},\hat{x}_t,\hat{y}_t}{p(\hat{x}_{t+1},\hat{x}_t,\hat{y}_t)\log_2{\frac{p(\hat{x}_{t+1},\hat{x}_t,\hat{y}_t)p(\hat{x}_t)}{p(\hat{x}_{t+1},\hat{x}_t)p(\hat{x}_t,\hat{y}_t)}}}.
\end{equation}
During this procedure, one need to determine the value of $q$. Marschinski and Kantz considered $q=2$ and $3$ in their research \cite{Marschinski-Kantz-2002-EPJB}, while Sandoval Jr. used $q=6$ and $24$ \cite{Sandoval-2014-Entropy}. In this work, we use $q=10$, $15$ and $20$ to calculate the transfer entropy between different sectors and present the results for $q=15$ since the results are similar.

\subsection{Information flow network}

For every pair of sectors $i$ and $j$, there are two quantities  $T^S_{i\rightarrow{j}}$ and $T^S_{j\rightarrow{i}}$ measuring the information interaction between them, where $T^S_{i\rightarrow{j}}$ measures the direction and strength of the information flow from sector $i$ to sector $j$ while $T^S_{j\rightarrow{i}}$ measures the direction and strength of the information flow from sector $j$ to sector $i$. We use the degree of asymmetric information flow (DAI or $\Delta{T}^S_{i\rightarrow{j}}$) introduced by Kwon and Oh to quantify the information effect between two stock sectors \cite{Kwon-Oh-2012-EPL}:
\begin{equation}
\Delta{T}^S_{i\rightarrow{j}}=T^S_{i\rightarrow{j}}-T^S_{j\rightarrow{i}}.
\label{Eq:DT}
\end{equation}
Following the above definition, we construct transfer entropy based information flow networks of the Chinese stock market. For two sectors $i$ and $j$, if $\Delta{T}^S_{i\rightarrow{j}}>0$, there is a directed link from sector $i$ to sector $j$ and the weight of the link is $\Delta{T}^S_{i\rightarrow{j}}$, on the contrary, if $\Delta{T}^S_{i\rightarrow{j}}<0$ then there is a directed link from sector $j$ to sector $i$ and the weight of the link is $\Delta{T}^S_{j\rightarrow{i}}=-\Delta{T}^S_{i\rightarrow{j}}$. Note that $\Delta{T}^S_{i\rightarrow{j}}=0$ holds if and only if the return time series of sector $i$ is exactly the same as the return time series of sector $j$, which is impossible in reality.

\subsection{Maximum spanning arborescence}

Two maximum spanning arborescences $A_{\max}$ (outgoing MSA $A_{\mathrm{out}}$ and incoming MSA $A_{\mathrm{in}}$) can be extracted from an information flow network. The outgoing MSA $A_{\mathrm{out}}$ has a source node and each vertex has exactly one incoming edge. The incoming MSA $A_{\mathrm{in}}$ has a sink node and each vertex has exactly one outgoing edge. For a connected network $G=(V,E)$ with $N$ nodes, a spanning arborescence $A$ is a directed loop-free subgraph that connects every node in the network with $N-1$ edges. There can be a lot of spanning arborescences for any given directed network. Denoting $w(\vec{e})$ be the weight (or DAI in this work) of edge $\vec{e}$, the total weight of spanning arborescence $A$ is
\begin{equation}
w(A) = \sum_{\vec{e}\in{A}}w(\vec{e}).
\end{equation}
A maximum spanning arborescence is the spanning arborescence whose edges have the total maximum total weight:
\begin{equation}
w(A_{\max})=\max_{A\in{G}}w(A).
\end{equation}
This can be implemented with the Chu-Liu/Edmond algorithm \cite{Chu-Liu-1965-SciSin,Edmonds-1967-JRNBS}.

\section{Data description}
\label{S1:Data}

The sector indices data used in this study is retrieved from Shenyin \& Wanguo Securities Co., Ltd., which are publicly available at http://www.swsresearch.com. In total, we have 28 sector indices of the Chinese stock market, which covers 3508 individual stocks. Each sector index ranges from 4 January 2000 to 29 December 2017, containing 4359 daily closing prices. The sector names and their corresponding six-digit codes are:
Agriculture \& forestry (801010), Automobile (801880), Bank (801780), Building \& decoration (801720), Building materials (801710), Chemical (801030), Commercial trade (801200), Communications (801770), Composite (801230), Computer (801750), Electrical equipment (801730), Electronic (801080), Food \& drink (801120), Household appliances (801110), Leisure \& services (801210), Light manufacturing (801140), Mechanical equipment (801890), Media (801760), Medicinal organisms (801150), Mining (801020), National defense (801740), Non-bank financial (801790), Non-ferrous metals (801050), Real estate (801180), Steel (801040), Textile and apparel (801130), Transportation (801170), and Utilities (801160).

\begin{table}[!h]
	\renewcommand\arraystretch{1}
	\centering
	\caption{Summary statistics of the return time series of the 28 SWS industry sectors from 4 January 2000 to 29 December 2017. To simplify the symbol, we use the last three digits of each 6-digit code to represent the corresponding Chinese stock market sector.}\tiny
	\begin{tabular}{p{0.25cm}<{\centering}p{1.7cm}<{\centering}p{0.4cm}<{\centering}p{0.3cm}<{\centering}p{0.5cm}<{\centering}p{0.3cm}<{\centering}p{0.5cm}<{\centering}p{0.4cm}<{\centering}p{0.5cm}<{\centering}}
		\hline\hline
		Symbol & Sector & Mean $10^{-3}$ & Max & Min & Std & Skewness & Kurtosis & JB  \\\hline 
		010&Agriculture \& forestry&0.241&0.092&-0.096& 0.020&-0.490& 6.202&2035.7\\
		020&Mining&0.275&0.095&-0.103& 0.021&-0.114& 5.840&1474.3\\
		030&Chemical&0.260&0.093&-0.092& 0.018&-0.522& 6.394&2289.8\\
		040&Steel&0.254&0.094&-0.097& 0.020&-0.296& 6.400&2162.5\\
		050&Non-ferrous metals&0.320&0.095&-0.101& 0.022&-0.326& 5.586&1291.5\\
		080&Electronic&0.276&0.094&-0.095& 0.021&-0.567& 5.512&1379.1\\
		110&Hosehold appliances&0.446&0.094&-0.094& 0.019&-0.253& 6.084&1773.7\\
		120&Food \& drink&0.555&0.092&-0.093& 0.017&-0.161& 6.469&2204.2\\
		130&Textile and apparel&0.225&0.093&-0.097& 0.019&-0.702& 6.940&3177.6\\
		140&Light manufacturing&0.245&0.094&-0.100& 0.019&-0.682& 6.931&3143.4\\
		150&Medicinal organisms&0.473&0.091&-0.091& 0.018&-0.485& 6.599&2522.9\\
		160&Utilities&0.221&0.095&-0.098& 0.017&-0.520& 7.205&3407.3\\
		170&Transportation&0.239&0.095&-0.101& 0.018&-0.488& 7.302&3532.9\\
		180&Real estate&0.358&0.094&-0.098& 0.020&-0.372& 5.929&1658.1\\
		200&Commercial trade&0.333&0.093&-0.097& 0.018&-0.546& 6.479&2414.1\\
		210&Leisure \& services&0.382&0.093&-0.098& 0.020&-0.451& 6.187&1992.0\\
		230&Composite&0.227&0.092&-0.097& 0.020&-0.660& 6.017&1968.7\\
		710&Bulding materials&0.406&0.090&-0.100& 0.020&-0.482& 6.130&1947.0\\
		720&Bulding \& decoration&0.247&0.094&-0.096& 0.019&-0.336& 6.488&2290.8\\
		730&Electrical equipment&0.381&0.095&-0.092& 0.019&-0.438& 5.926&1693.9\\
		740&National defense&0.323&0.096&-0.102& 0.023&-0.342& 5.919&1631.6\\
		750&Computer&0.322&0.095&-0.101& 0.021&-0.358& 5.361&1105.2\\
		760&Media&0.304&0.095&-0.105& 0.022&-0.368& 5.340&1092.9\\
		770&Communications&0.219&0.095&-0.099& 0.020&-0.215& 6.125&1807.1\\
		780&Bank&0.287&0.096&-0.105& 0.019& \textbf{0.182}& 7.128&3117.5\\
		790&Non-bank financial&0.341&0.095&-0.102& 0.024& \textbf{0.041}& 5.523&1157.4\\
		880&Automobile&0.383&0.093&-0.098& 0.020&-0.449& 6.236&2048.8\\
		890&Mechanical equipment&0.381&0.093&-0.093& 0.019&-0.527& 6.292&2169.6\\
		\hline\hline
		\label{TB:Ret:Stat}
	\end{tabular}
\end{table}

Table~\ref{TB:Ret:Stat} presents the summary statistics of the return time series. We can see that the mean of the logarithmic returns of all the sectors are positive ranging from 0.000219 to 0.000555, which means that the Chinese stock market exhibits a slow overall rising trend during the sample period. All of the returns varies from -10\% to 10\% constrained by the $\pm10\%$ price limit rule in the Chinese stock market, where some values exceed the $\pm10\%$ range due to the tick size of one cent. According to column of standard deviation, the \textit{non-bank financial} sector fluctuates the most, while the \textit{food \& drink} sector is the least volatile. It means that the {\textit{non-bank financial}} sector is riskier than the \textit{food \& drink} sector in the Chinese stock market. All the return distributions are left-skewed except for the {\textit{bank}} sector ($\mathrm{skewness}=0.182$) and the {\textit{non-bank financial}} sector ($\mathrm{skewness}=0.041$). We assume that these two finance-related sectors have a special economic position in the Chinese stock market and we will give some insights in our empirical results. The kurtosis of each return time series is positive, means that the overall return distribution is broader than the normal distribution, which is known to be ``fat-tailed'' or ``heavy-tailed'', a well-established stylized fact of financial returns \cite{Gopikrishnan-Meyer-Amaral-Stanley-1998-EPJB,Gabaix-Gopikrishnan-Plerou-Stanley-2003-Nature}. According to the Jarque-Bera test, we find that all the Jarque-Bera test statistics are much greater than the critical value 9.442 with the significant level $\alpha=0.01$. Hence the return distributions of all the 28 sectors are not normally distributed.

\section{Empirical results}
\label{S1:Results}

\subsection{Maximum spanning arborescences of the whole sample}

\Figure[htbp!](topskip=0pt, botskip=0pt, midskip=0pt){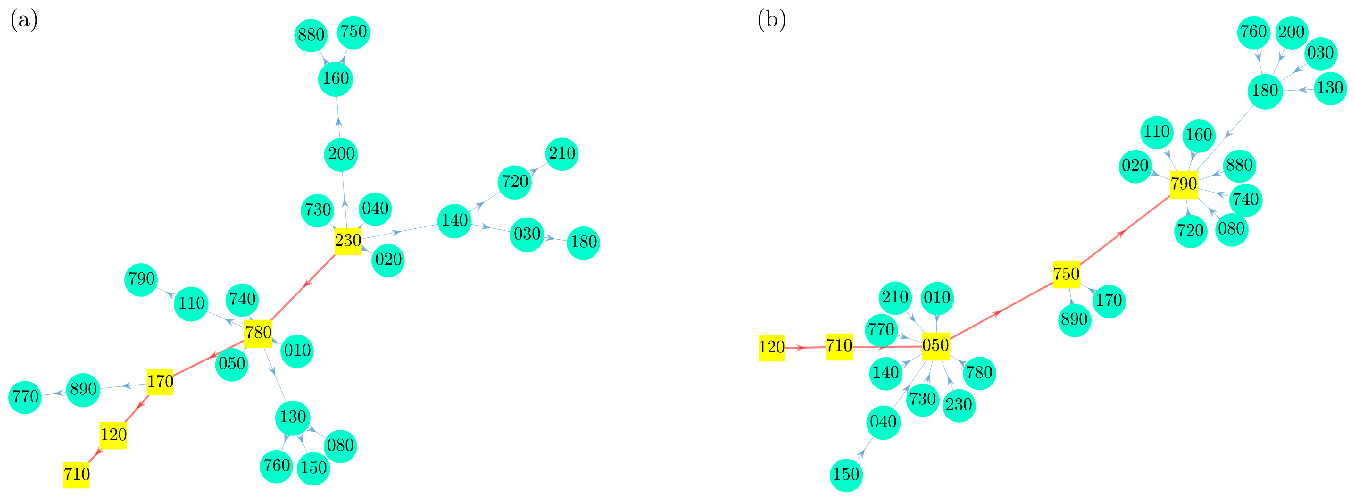}
{The maximum spanning arborescences of the information flow network among the 28 Chinese stock market sectors from 2000 to 2017 and the maximal information flow path (we use red edge with yellow square markers to make a distinction). (a) Outgoing MSA. (b) Incoming MSA.\label{Fig:MST:OI:whole}}

As mentioned in Section \ref{S1:Method}, transfer entropy can proxy for the strength and direction of the information flow between two time series. Following Oh et al. \cite{Oh-Oh-Kim-Kwon-2014-JKPS}, we use the degree of asymmetric information flow (DAI) to measure the net information flow between stock sectors and construct directed information flow networks. We advance their method by identifying the maximum spanning arborescences by connecting the highest DAI of each network and the maximal information flow path which has the highest DAI from the starting node (no inflow from others) to the ending node (no outflow to others). The outgoing and incoming MSAs of the 28 Chinese stock sectors from 2000 to 2017 are illustrated in Fig.~\ref{Fig:MST:OI:whole}.

According to Fig.~\ref{Fig:MST:OI:whole}(a), the \textit{composite} sector (code 230) locates at the root of the outgoing MSA, which acts as the information source. It implied that the \textit{composite} sector is the most influencing sector. The maximal information flow path travels from the \textit{composite} sector to the \textit{building materials} sector through the \textit{bank} sector (code 780), the \textit{transportation} sector (code 170), and the \textit{food \& drink} sector (code 120), along which the \textit{bank} sector (code 780) plays a crucial role because its outgoing degree 6 is the highest and thus it has a direct information impact on other six sectors. The important role of the \textit{composite} sector as an information source might stem from the fact that this sector is closely related to our daily life and feels rapidly the change in sentiment.

From Fig.~\ref{Fig:MST:OI:whole}(b), the \textit{non-bank financial} sector (code 790) lies at the center of the incoming MSA, acting as the information sink of the information flow network. It means that the \textit{non-bank financial} sector is the most influenced sector. The maximal information flow path of the incoming MSA starts from the \textit{food \& drink} sector (code 120), passes through the \textit{building materials} (code 710) sector, the \textit{non-ferrous metals} sector (code 050) and the \textit{computer} sector (code 750), and finally ends at the \textit{non-bank financial} sector. Along path, the \textit{non-ferrous metals} sector (code 050) is also a hub, affected by other four sectors including the \textit{media} sector (code 760), the \textit{commercial trade} sector (code 200), the \textit{chemical} sector (code 030) and the \textit{textile and apparel} sector (code 130). The observation that the \textit{non-bank financial} sector acts as a main information sink can be partly explained by the fact that this sector is a main investment channel for the Chinese people. The fluctuation in other sectors impact the savings of people and their investment decisions among real-estates, bank savings, stocks and non-bank financial products.

\subsection{Yearly evolution of the maximum spanning arborescences}

\begin{figure*}[!htb]
	\centering
	\includegraphics[width=\textwidth]{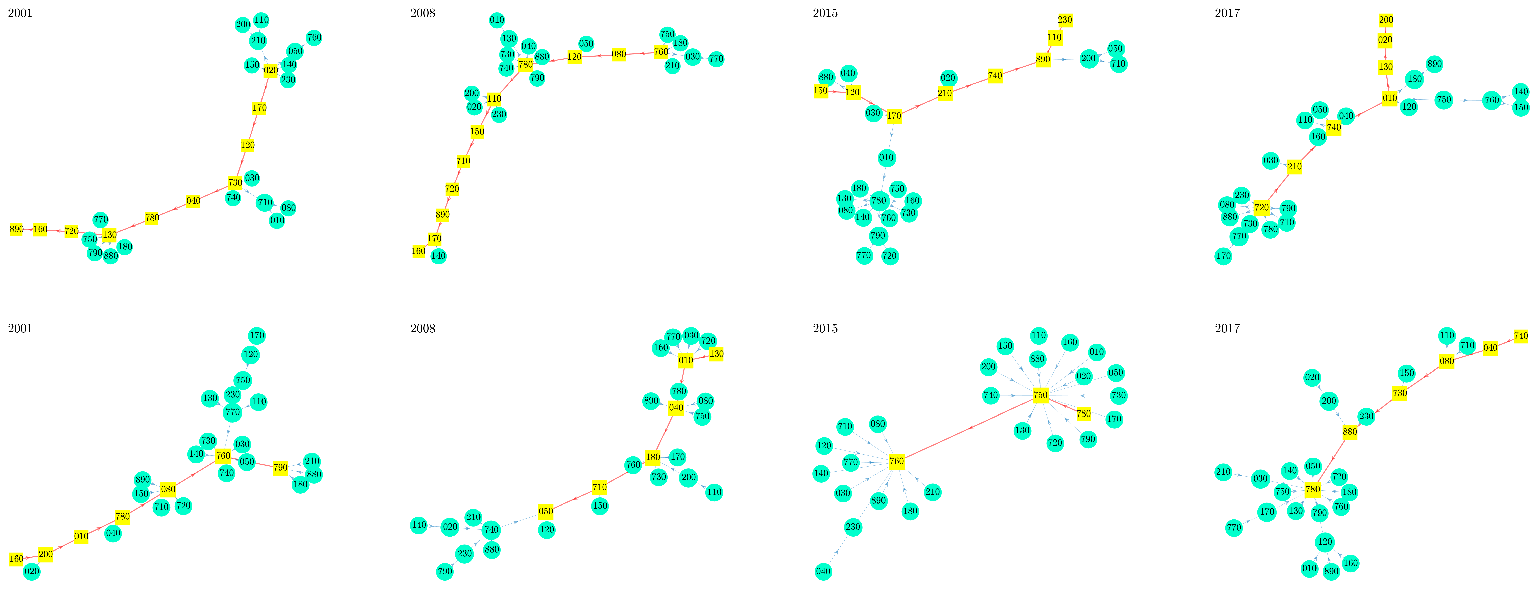}
	\caption{Outgoing (top row) and incoming (bottom row) maximum spanning arborescences and the correspond maximum information flow paths between 28 Chinese stock market sectors in 2001, 2008, 2015 and 2017.}
	\label{Fig:4y}
\end{figure*}
Different MSAs have different patterns indicating that there are different hierarchical structures of information flow. Yao et al. studied information transfer routes between cross-industry and cross-region electricity consumption data based on transfer entropy and MST and found that the MSTs follow a chain-like structure in developed provinces and star-like structures in developing provinces \cite{Yao-Kuang-Lin-Sun-2017-Entropy}. Following their study, we investigate evolving shapes of the yearly MSAs. Figure~\ref{Fig:4y} shows the yearly outgoing and incoming MSAs in four representative years, together with the correspond maximal information flow paths. It can be seen that the shapes and the root nodes of the MSAs change with time. 

The maximum information flow path on the outgoing MSA in 2001 is $020\to 170 \to 120 \to 730 \to 040 \to 780 \to 130 \to 720 \to 160 \to 890$. It contains 10 sectors and its total information is 58.27.
The maximum information flow path on the outgoing MSA in 2008 is $760 \to 080 \to 120 \to 780 \to 110 \to 150 \to 710 \to 720 \to 890 \to 170 \to 160$, which has 11 sectors and total information of 85.15.
The maximum information flow path on the outgoing MSA in 2015 is $150 \to 120 \to 170 \to 210 \to 740 \to 890 \to 110 \to 230$, which possesses 8 sectors with the total information flow being 67.35.
The maximum information flow path on the outgoing MSA in 2017 is $720 \to 210 \to 740 \to 010 \to 130 \to 020 \to 200$, which consists of 7 sectors with the total information flow equal to 57.76.
These paths changed a lot in different years. The situation is very similar for the incoming MSAs. 
The information about the root node sectors and the maximal information paths are presented in Table~\ref{TB:Omst} for the yearly outgoing MSAs and in Table~\ref{TB:Imst}  for the yearly incoming MSAs for each year.
\begin{table}
	\renewcommand\arraystretch{1.5}
	\centering
	\caption{\textbf{The evolution of the maximal information outflow path of the outgoing MSAs.}}
	\smallskip
	\tiny
	\begin{tabular}{p{0.3cm}<{\centering}p{0.5cm}<{\centering}p{5cm}<{\centering}p{0.5cm}<{\centering}p{0.4cm}<{\centering}}
		\hline\hline
		\textbf{Year} & \textbf{Root sector} & \textbf{Maximal information outflow path} & \textbf{No. of sectors} & \textbf{DAI ($10^{-2})$}\\
		\hline
		2000&780&780$\rightarrow$110$\rightarrow$890$\rightarrow$750$\rightarrow$230$\rightarrow$150&6&39.95\\
		2001&020&020$\rightarrow$170$\rightarrow$120$\rightarrow$730$\rightarrow$040$\rightarrow$780$\rightarrow$130$\rightarrow$720$\rightarrow$160$\rightarrow$890&10&58.27\\
		2002&180&180$\rightarrow$200$\rightarrow$740$\rightarrow$020$\rightarrow$010$\rightarrow$230$\rightarrow$040$\rightarrow$790$\rightarrow$750$\rightarrow$030$\rightarrow$050&11&70.11\\
		2003&790&790$\rightarrow$150$\rightarrow$030$\rightarrow$750$\rightarrow$110$\rightarrow$040$\rightarrow$230$\rightarrow$710&8&55.82\\
		2004&750&750$\rightarrow$150$\rightarrow$790$\rightarrow$010$\rightarrow$130$\rightarrow$040&6&62.33\\
		2005&790&790$\rightarrow$200$\rightarrow$040$\rightarrow$170$\rightarrow$150$\rightarrow$160$\rightarrow$760$\rightarrow$210$\rightarrow$770$\rightarrow$730&10&86.13\\
		2006&080&080$\rightarrow$010$\rightarrow$120$\rightarrow$740&4&42.9\\
		2007&200&200$\rightarrow$080$\rightarrow$150$\rightarrow$160$\rightarrow$750$\rightarrow$010$\rightarrow$050&7&70.89\\
		2008&760&760$\rightarrow$080$\rightarrow$120$\rightarrow$780$\rightarrow$110$\rightarrow$150$\rightarrow$710$\rightarrow$720$\rightarrow$890$\rightarrow$170$\rightarrow$160&11&85.15\\
		2009&790&790$\rightarrow$040$\rightarrow$230$\rightarrow$770$\rightarrow$020$\rightarrow$760$\rightarrow$880&7&56.76\\
		2010&780&780$\rightarrow$180$\rightarrow$040$\rightarrow$020$\rightarrow$010$\rightarrow$120&6&61.34\\
		2011&760&760$\rightarrow$160$\rightarrow$740$\rightarrow$170$\rightarrow$230$\rightarrow$130$\rightarrow$140&7&75.37\\
		2012&780&780$\rightarrow$750$\rightarrow$230$\rightarrow$010$\rightarrow$120$\rightarrow$210$\rightarrow$790$\rightarrow$020$\rightarrow$150&9&91.48\\
		2013&050&050$\rightarrow$790$\rightarrow$020$\rightarrow$130$\rightarrow$710$\rightarrow$080&6&51.18\\
		2014&160&160$\rightarrow$780$\rightarrow$040$\rightarrow$790$\rightarrow$150&5&39.41\\
		2015&150&150$\rightarrow$120$\rightarrow$170$\rightarrow$210$\rightarrow$740$\rightarrow$890$\rightarrow$110$\rightarrow$230&8&67.35\\
		2016&160&160$\rightarrow$040$\rightarrow$180$\rightarrow$050$\rightarrow$750$\rightarrow$010&6&44.04\\
		2017&720&720$\rightarrow$210$\rightarrow$740$\rightarrow$010$\rightarrow$130$\rightarrow$020$\rightarrow$200&7&57.76\\
		\hline\hline
	\end{tabular}
	\label{TB:Omst}
\end{table}

\begin{table}
	\renewcommand\arraystretch{1.5}
	\centering
	\caption{\textbf{The evolution of the maximal information inflow path of the incoming MSAs.}}
	\smallskip
	\tiny
	\begin{tabular}{p{0.3cm}<{\centering}p{0.5cm}<{\centering}p{5cm}<{\centering}p{0.5cm}<{\centering}p{0.4cm}<{\centering}}
		\hline\hline
		\textbf{Year} & \textbf{Root sector} & \textbf{Maximal information inflow path} & \textbf{No. of sectors} & \textbf{DAI ($10^{-2}$)}\\\hline
		2000&040&780$\rightarrow$740$\rightarrow$710$\rightarrow$020$\rightarrow$040&5&45.16\\
		2001&790&160$\rightarrow$200$\rightarrow$010$\rightarrow$780$\rightarrow$080$\rightarrow$760$\rightarrow$790&7&40.29\\
		2002&780&770$\rightarrow$160$\rightarrow$020$\rightarrow$730$\rightarrow$780&5&26.31\\
		2003&130&230$\rightarrow$730$\rightarrow$720$\rightarrow$710$\rightarrow$180$\rightarrow$210$\rightarrow$020$\rightarrow$890$\rightarrow$130&9&78.82\\
		2004&740&790$\rightarrow$010$\rightarrow$720$\rightarrow$760$\rightarrow$230$\rightarrow$200$\rightarrow$020$\rightarrow$740&8&71.2\\
		2005&750&160$\rightarrow$760$\rightarrow$230$\rightarrow$710$\rightarrow$180$\rightarrow$880$\rightarrow$080$\rightarrow$750&8&72.2\\
		2006&050&120$\rightarrow$740$\rightarrow$890$\rightarrow$210$\rightarrow$780$\rightarrow$050&6&50.09\\
		2007&130&140$\rightarrow$030$\rightarrow$120$\rightarrow$180$\rightarrow$710$\rightarrow$020$\rightarrow$050$\rightarrow$130&8&70.69\\
		2008&050&130$\rightarrow$010$\rightarrow$040$\rightarrow$180$\rightarrow$710$\rightarrow$050&6&54.56\\
		2009&710&180$\rightarrow$150$\rightarrow$730$\rightarrow$030$\rightarrow$210$\rightarrow$740$\rightarrow$780$\rightarrow$710&8&73.42\\
		2010&030&180$\rightarrow$760$\rightarrow$200$\rightarrow$730$\rightarrow$120$\rightarrow$040$\rightarrow$030&7&67.84\\
		2011&120&780$\rightarrow$730$\rightarrow$200$\rightarrow$050$\rightarrow$880$\rightarrow$770$\rightarrow$040$\rightarrow$120&8&51.71\\
		2012&180&160$\rightarrow$740$\rightarrow$110$\rightarrow$200$\rightarrow$180&5&46.73\\
		2013&150&890$\rightarrow$760$\rightarrow$210$\rightarrow$080$\rightarrow$150&5&46.29\\
		2014&110&790$\rightarrow$150$\rightarrow$050$\rightarrow$210$\rightarrow$120$\rightarrow$110&6&60.12\\
		2015&760&780$\rightarrow$750$\rightarrow$760&3&40.05\\
		2016&230&170$\rightarrow$710$\rightarrow$110$\rightarrow$760$\rightarrow$200$\rightarrow$010$\rightarrow$770$\rightarrow$030$\rightarrow$140$\rightarrow$080$\rightarrow$230&11&75.17\\
		2017&780&740$\rightarrow$040$\rightarrow$080$\rightarrow$730$\rightarrow$880$\rightarrow$780&6&59.92\\
		\hline\hline
	\end{tabular}
	\label{TB:Imst}
\end{table}

\begin{figure*}[!htb]
	\centering
	\includegraphics[width=15cm]{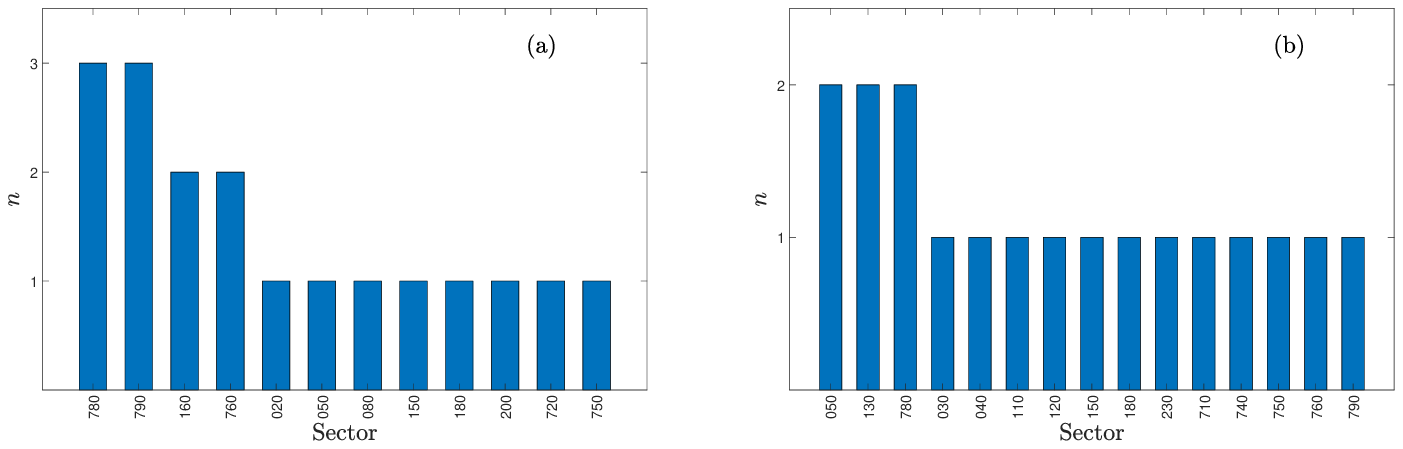} 
	\caption{Occurrence numbers $n$ of sectors as a source in the yearly outgoing MSAs (a) and as a sink in the yearly incoming MSAs (b).}
	\label{Fig:root:bar}
\end{figure*}

The root nodes play an important role in the corresponding maximum spanning arborescences. The occurrence numbers of sectors as a source in the outgoing MSAs and as a sink in the incoming MSAs are shown in Fig.~\ref{Fig:root:bar}. It shows that the \textit{bank} and \textit{non-bank finance} sectors appeared  three times as the source node in the outgoing MSAs. Net information transfers from these two sectors to other sectors, indicating the importance of financial firms in delivering market information in the the Chinese market. Secondly, the \textit{public utilities} and \textit{media} sectors appeared twice, whereas the \textit{excavation}, \textit{non-ferrous metals}, \textit{electronics}, \textit{medical biology}, \textit{real estate}, \textit{commercial digital}, \textit{architectural decoration} and \textit{Computers} sectors appeared once, indicating that these sectors have information transmission to the entire market and affect the structure of the market. The variation of the maximum paths and the source sector also implies the presence of the section rotation phenomenon \cite{Grauer-Hakansson-Shen-1990-JBF,Grauer-Hakansson-2001-RQFA,Leibon-Pauls-Rockmore-Savell-2008-PNAS}.
On the other hand, it can be seen that there are three sectors (\textit{non-ferrous metals}, \textit{textiles and apparel} and \textit{bank}) appeared twice as the sink nodes in the incoming MSA, showing that these three sectors receive the information flow from other sectors and become more representative among the whole market. In addition, the \textit{chemical}, \textit{steel}, \textit{household appliances}, \textit{food and beverage}, \textit{medical biology}, \textit{real estate}, \textit{comprehensive}, \textit{building materials}, \textit{national defense military}, \textit{computer}, \textit{media} and \textit{non-bank finance} sectors appeared once as the sink node in the incoming MSAs, meaning that these sectors are also quite information-sensitive in different years.

As shown in Fig.~\ref{Fig:MST:OI:whole}, there are usually several hubs with relatively high degrees in the outgoing and incoming MSAs, which can be viewed as secondary sources and sinks.
We identify from the 36 MSAs seven sectors (\textit{non-bank finance}, \textit{bank}, \textit{computer}, \textit{media}, \textit{real estate}, \textit{medical biology} and \textit{non-ferrous metals}), which are secondary information sources and/or sinks. Consistent with our common intuition, these sectors are prone to influence other sectors or be influenced by other sectors. The \textit{bank} sector has the highest degree of the information outflow and inflow as the root node, which means that it occupies a very important position in the information flow networks of the Chinese stock market. It is because that the banking system is often adopted as an important tool for the Chinese government to regulate the Chinese financial markets and other industries.
\begin{figure*}[!htbp]
	\centering
	\includegraphics[width=13cm]{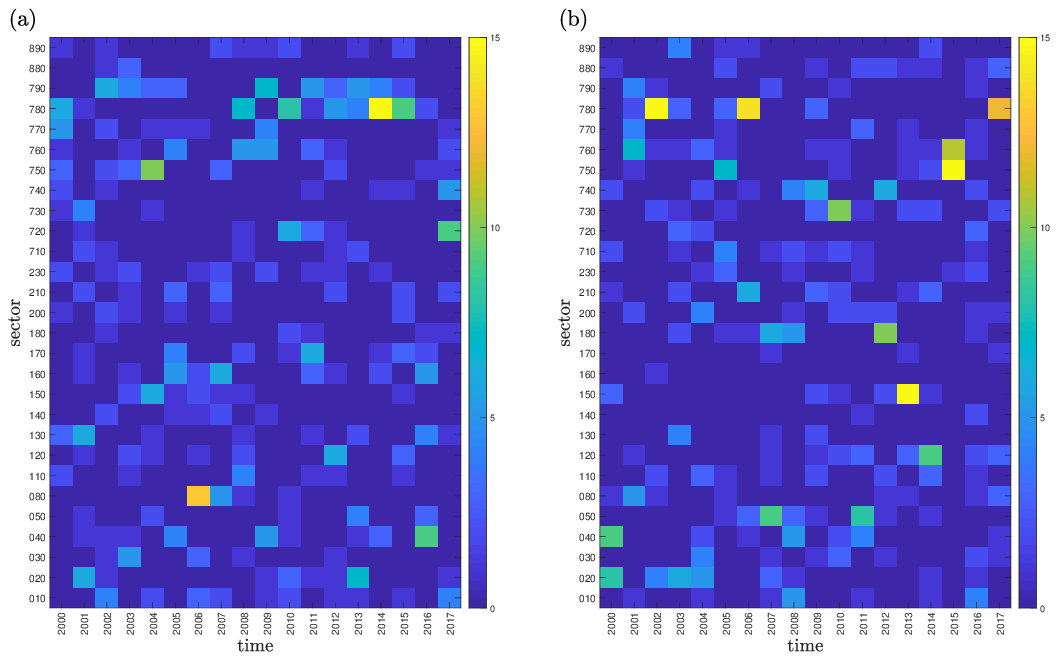}
	\caption{Heat maps of the degree of different sectors in different information flow networks from 2000 to 2017. (a) outgoing information flow networks. (b) incoming information flow networks. A higher value of degree will be reflected by a brighter color in our heat map.}
	\label{Fig:mst:d:t}
\end{figure*}

Sectors with high degrees in the MSAs play an important role in the information flow networks. For a given outgoing information flow network, the higher the degree of the sector, the more sectors are directly affected by it in terms of the information interaction. On the contrary, for a given incoming information flow network, the higher the degree of the sector, the sector receives information from more sectors, and more sectors directly affect the sector. We use two heat maps to show the degree of different sectors in the outgoing and incoming MSAs in Fig.~\ref{Fig:mst:d:t}. 
Our analysis also shows that the outdegree and indegree of each sector in the information flow networks vary with time, it means that the impact of various sectors on the stock market varies from time to time. It reflects a well known phenomenon of ``industry rotation'' or ``sector rotation'' in the stock market all over the world \cite{Grauer-Hakansson-Shen-1990-JBF,Grauer-Hakansson-2001-RQFA,Leibon-Pauls-Rockmore-Savell-2008-PNAS}. As it can be clearly seen from Fig.~\ref{Fig:mst:d:t}(a), the \textit{non-bank finance} and \textit{bank} sectors have significant high degrees in the outgoing MSAs after 2008. We can conclude that these two sectors are the key sectors in the economic recovery stage, proxied by the five trillion yuan bailout action in China. In Fig.~\ref{Fig:mst:d:t}(b), we can see that the \textit{bank} sector has very high degree in 2002, 2006 and 2017, it reflects the fact that the \textit{bank} sector is a very special sector compared with others in the Chinese stock market. Another very interesting result is that there is a double center-like information flow structure based on the \textit{media} and \textit{computer} sectors in 2015 which means these two sectors have most information interaction with other sectors and become the most representative ones in that year. Indeed, the stock market bubble in 2015 was fueled by the ``mass entrepreneurship and innovation'' policy of the central government, during which a huge number of startup enterprises in the ``Internet+'' fields were registered.

\subsection{The MSAs before, during and after stock market turmoils}

\begin{figure}[!htp]
	\centering
	\includegraphics[width=9cm]{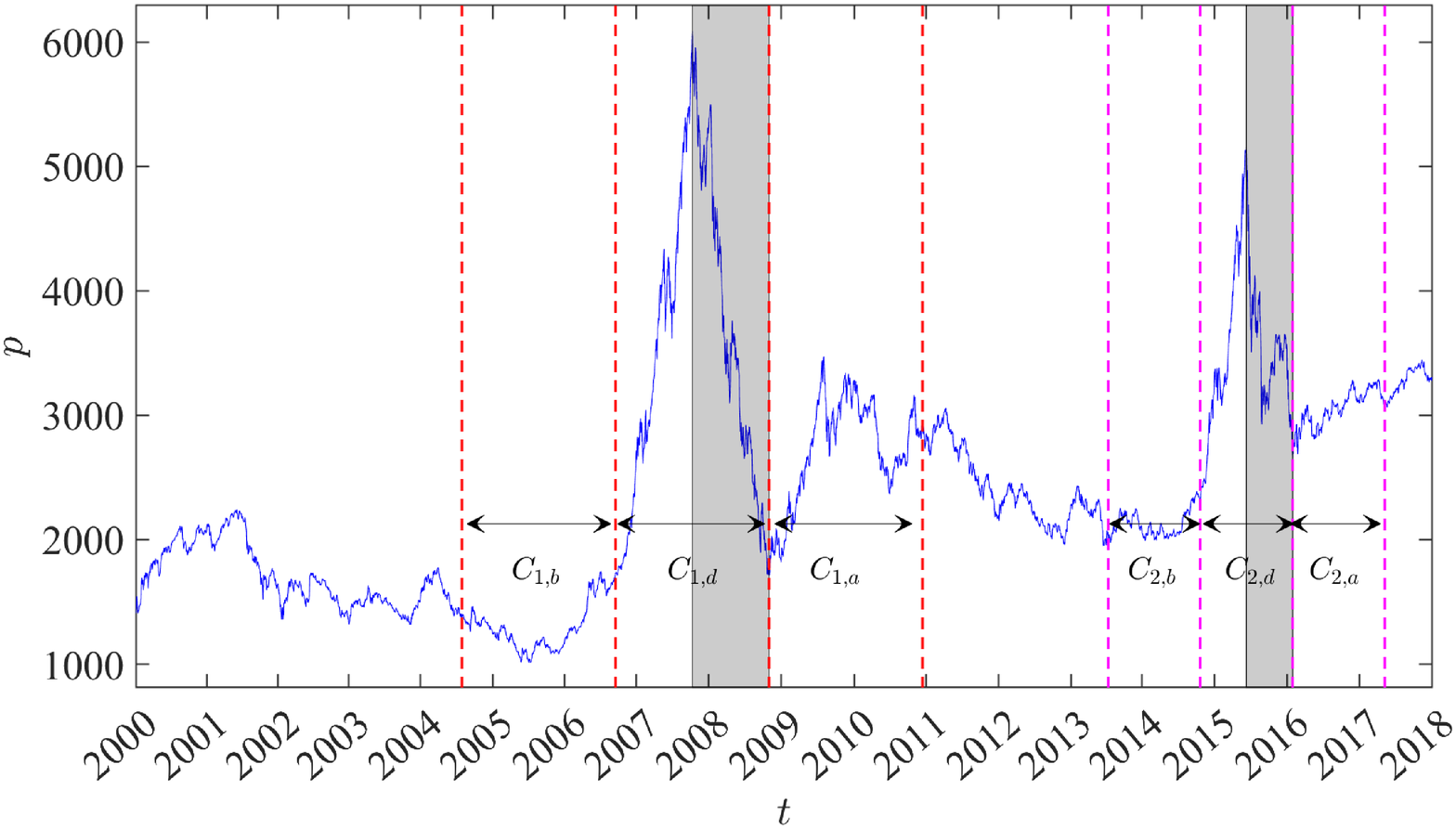}
	\caption{The closing price of Shanghai Stock Exchange Composite Stock Price (SSEC) Index from 04 January 2000 to 29 December 2017. The two gray regions are on behalf of two crashes of the Chinese stock market which from 16 October 2007 to 04 November 2008 and 12 June 2015 to 28 January 2016 respectively.}
	\label{Fig:Crash}
\end{figure}

The structure of information flow network usually changes a lot around financial turmoils (see \cite{Han-Xie-Xiong-Zhang-Zhou-2017-FNL} and references therein). For instance, Jang et al. studied the impact of currency crises on the MST structure of stock markets \cite{Jang-Lee-Chang-2011-PA}. Motivated by this, we investigate the evolution of the MSA structure of the Chinese stock market before, during and after two stock market turmoils. As shown in Fig.~\ref{Fig:Crash}, the two crashes range respectively from 16 October 2007 to 4 November 2008 spanning 259 trading days and from 12 June 2015 to 28 January 2016 spanning 155 trading days. We use the data of the same length of trading days before and after the starting date of a crash as the data sample during the market turmoils (a bubble followed by a crash), denoted as $C_{1,d}$ and $C_{2,d}$ in Fig.~\ref{Fig:Crash}. In order to analyze and compare the changes of MSTs of the information flow networks, the same time span data before and after the turmoil periods are used as the control data samples, which are denoted as $C_{1,b}$ and $C_{1,a}$ for the first case and $C_{2,b}$ and $C_{2,a}$ for the second case.

\begin{figure*}[!htp]
	\centering
	\includegraphics[width=15cm]{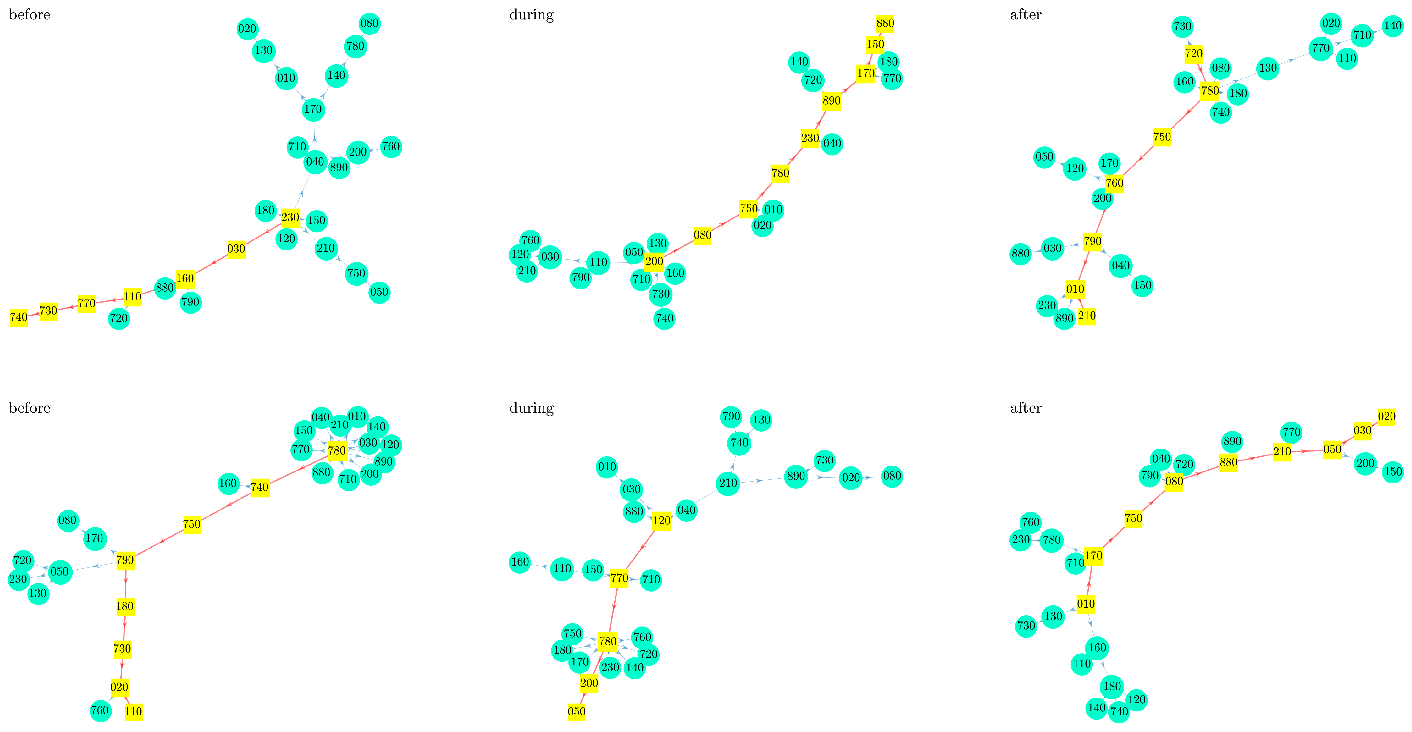}
	\caption{The outgoing MSAs before (left column), during (middle column) and after (right column) the large stock market turmoils around the end of 2007 (top row) and around 2015 (bottom row).}
	\label{Fig:compcr:outflow}
\end{figure*}
\begin{figure*}[!htp]
	\centering
	\includegraphics[width=15cm]{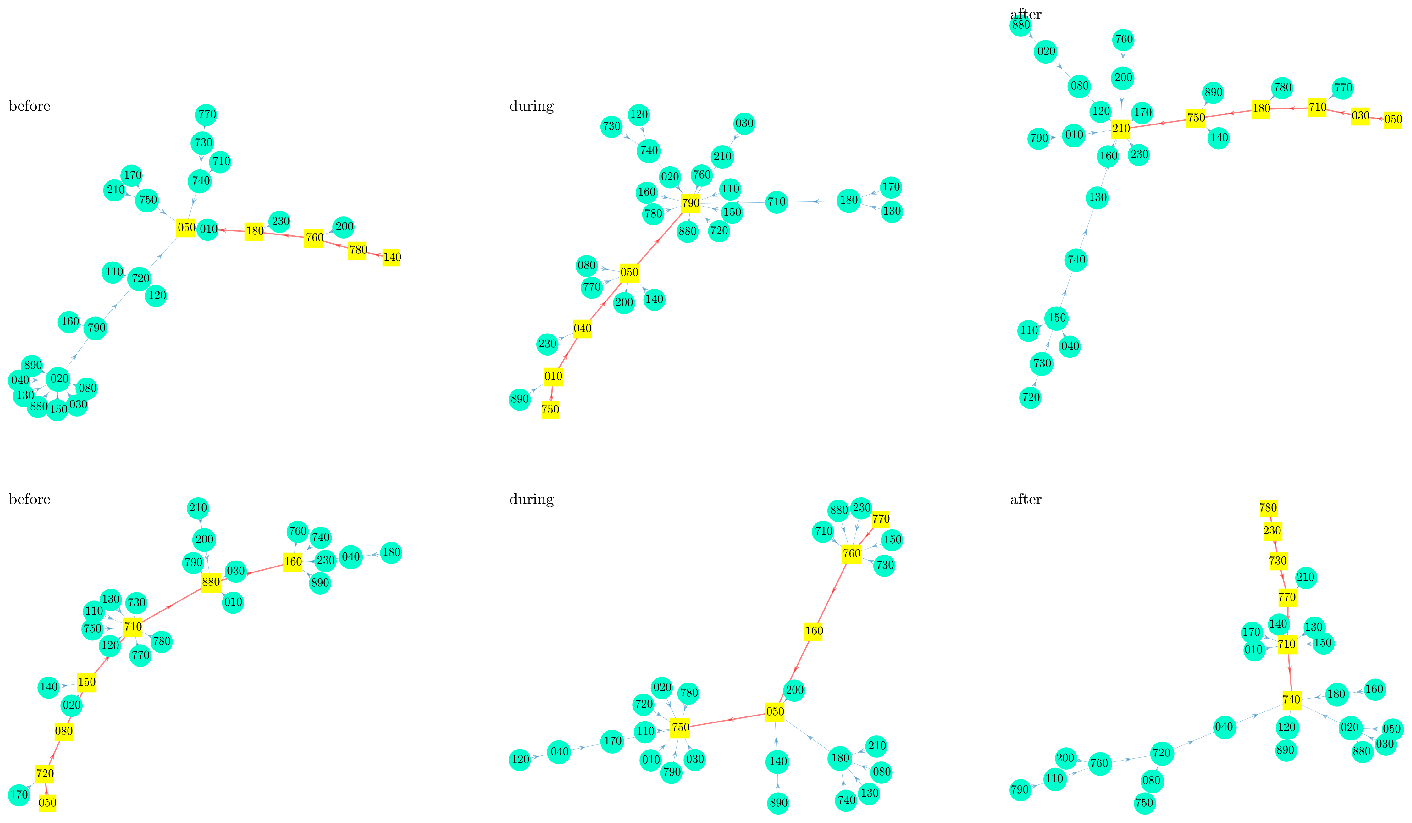}
	\caption{The incoming MSAs before (left column), during (middle column) and after (right column) the large stock market turmoils around the end of 2007 (top row) and around 2015 (bottom row).}
	\label{Fig:compcr:inflow}
\end{figure*}

We extract the outgoing and incoming MSAs before, during and after the two large stock market turmoil periods, which are illustrated in Fig.~\ref{Fig:compcr:outflow} and Fig.~\ref{Fig:compcr:inflow}. Figure~\ref{Fig:compcr:outflow} shows that, before the two turmoils, the \textit{household appliances}, \textit{electrical equipment} and \textit{national defense} sectors appeared at the same time in the maximal information flow paths, which suggests that these sectors were more active in the transmission of information flow before market turmoil. During the market turmoils, both the \textit{commercial trade} and \textit{bank} sectors appeared in the paths, indicating that these two sectors played a major role in information transmission during market turmoil. After the market turmoils, the \textit{leisure \& services} and \textit{computer} sectors became more active in the information transfer. With respect to the incoming MSAs in Fig.~\ref{Fig:compcr:inflow}, before these two market turmoil periods, the \textit{non-ferrous metals} sector made more information transfer along the paths. The \textit{non-ferrous metals} and \textit{computer} sectors became more active during market turmoils, while the \textit{building materials} sector participated much information transfer after the market turmoils. These structural variations also signal the presence of sector rotation in the Chinese stock market in the sense that different sectors dominate in the information transmission process in the evolving market.

\begin{figure*}[!ht]
	\centering
	\includegraphics[width=13cm]{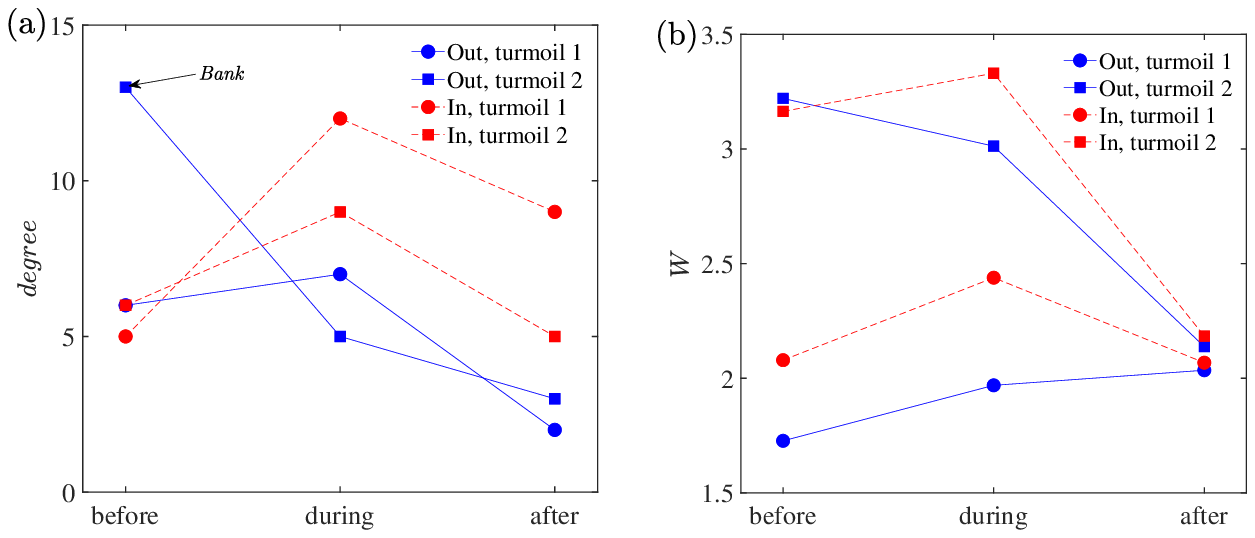}
	\caption{The degree of the root sectors (a) and the total weight of the maximal information flow path (b) of different MSTs before, during and after the two market turmoil periods.}
	\label{Fig:crash:D:W}
\end{figure*}

We further compare the degrees of the source and sink sectors and the total weights along the maximum paths of the outgoing and incoming MSAs, which are shown in Fig.~\ref{Fig:crash:D:W}. It can be seen clearly in Fig.~\ref{Fig:crash:D:W}(a) that the degree of the root sectors in all these four types MSAs rose first and then fell except for the outgoing MSA before the second market turmoil. The root of this outgoing  MSA is the \textit{bank} sector and there is a clearly cluster around it as shown in Fig.~\ref{Fig:compcr:outflow}. The \textit{bank} sector was very special because the bubble in 2015 was mainly fueled by the ``mass entrepreneurship and innovation'' policy during which investing in Internet+ startup firms was hot and by the high-leverage financing in stock trading. The higher degrees of the source and sink sectors during market turmoils imply that the information transfer behavior is more concentrated and the market is more synchronized when the market is very volatile. From Fig.~\ref{Fig:crash:D:W}(b), all the maximum information flow paths have similar total weight after market turmoils, while the information flows differed a lot before and during market turmoils.

\subsection{The specificity of root sectors of outgoing and incoming MSAs}

In order to study the particularity of the root node sectors in the information flow networks, we use the correlation coefficient between the yearly return of each sector and the yearly return of the Shanghai Stock Exchange Composite index price to measure the representative strength of the sector in the entire market. If the correlation coefficient between the sector and SSEC is larger, it indicates that the trend of the sector is more similar to the Shanghai Composite index and the sector is more representative of the market behavior. On the contrary, if the correlation coefficient is smaller, then the sector is less representative of the whole market.

The correlation coefficient between two time series $X_i$ and $Y_i$ ($i=1,\dots,n$) can be calculated as follows
\begin{equation}
\rho_{X,Y}=\dfrac{\sum_{i=1}^{n}(X_i-\bar{X})(Y_i-\bar{Y})}{\sqrt{\sum_{i=1}^{n}(X_i-\bar{X})^2}\sqrt{\sum_{i=1}^{n}(Y_i-\bar{Y})^2}}.
\label{Eq:corr}
\end{equation}
We first calculate the correlation coefficients between the return time series of different sector indices and the Shanghai Composite index and then calculate the correlation coefficients between the root sector indices of the outgoing and incoming information flow MSAs and the Shanghai Composite index. For each root node sector, we investigate the correlation coefficient $\rho(r_{\mathrm{root}}, r_{\mathrm{ssec}})$, where $r_{\mathrm{root}}$ is the yearly return time series of the root sector and $r_{\mathrm{ssec}}$ is the yearly return series of the Shanghai Composite index. Note that the root sectors are generally not the same in different years. As a control group, for each year, we randomly select the yearly return time series of non-root sectors to form a control sample $r_t^s$ ($t=2000, \ldots, 2017$) and calculate the correlation coefficients between these randomly selected sectors and the Shanghai Composite index. 

\begin{figure}[!htbp]
	\centering
	\includegraphics[width=9cm]{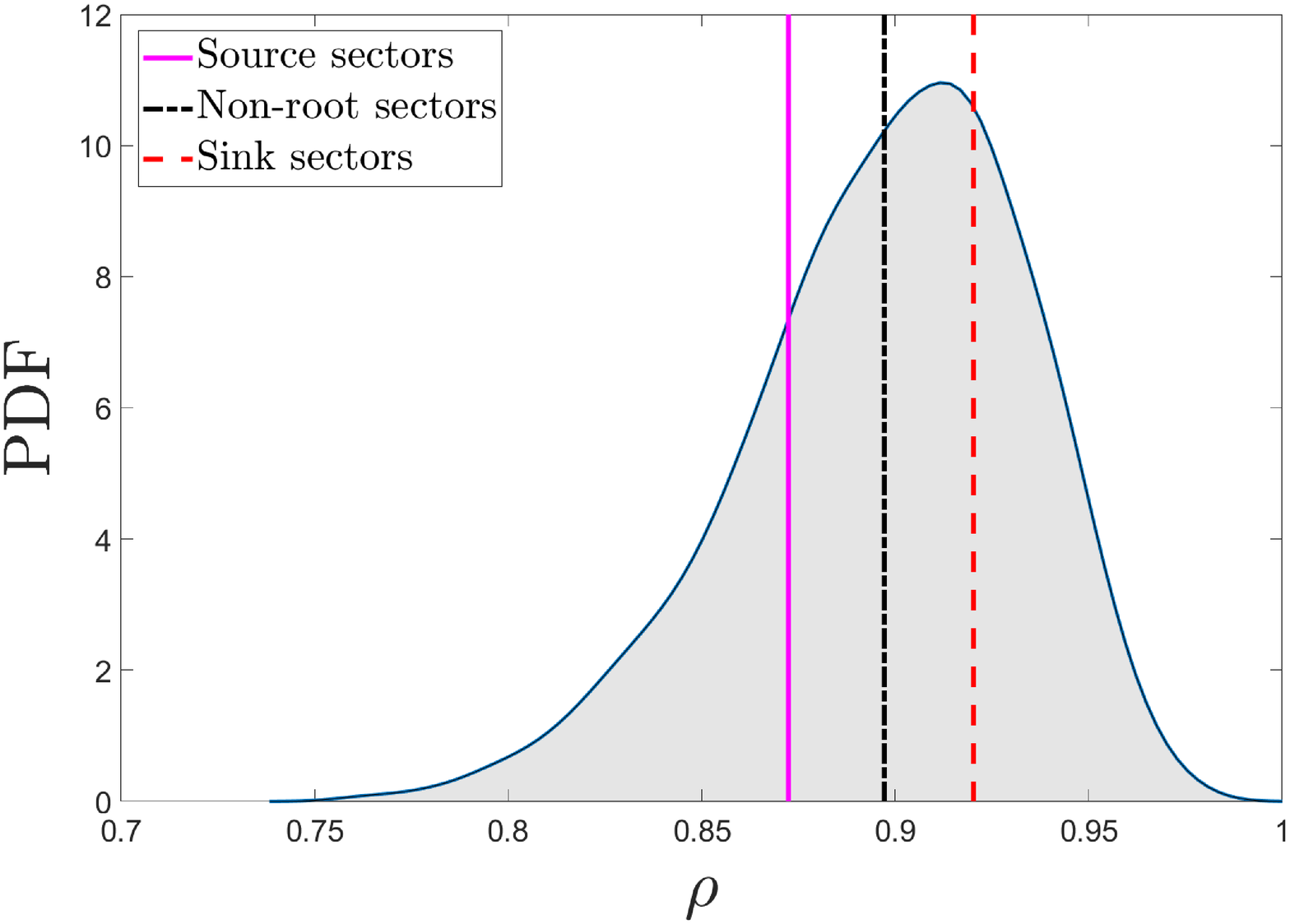}
	\caption{Comparison of the correlation coefficients between the returns of different types of sectors (source, sink and non-root) and the SSEC index.}
	\label{Fig:ret:root}
\end{figure}

We use a PDF to show the result of the correlation coefficients between the return time series of the randomly selected non-root sectors and the SSEC in Fig.~\ref{Fig:ret:root}. It shows that the correlation coefficient between non-root sectors and the Shanghai Composite Index fluctuates around 0.9 (the black vertical line), indicating that there is a high correlation between different sector indices and the Shanghai Composite Index, which verifies the overall consistency of the Chinese stock market. As for the root sectors of the outgoing and incoming information flow MSAs, the correlation coefficients deviate from the mean correlation coefficient for the non-root sectors, which provides strong evidence for the particularity of the root sectors. With respect to the source sector of the outgoing information flow MSAs, the correlation coefficient (0.8724) is smaller than the mean value of the control group, which indicates that source sectors possess or produce idiosyncratic information and exhibit active traits. On the other hand, the sink sectors of the incoming information flow MSAs have a significant high correlation coefficient (0.9202), which indicates they exhibit passive traits and are more representative of the Chinese stock market.

\section{Conclusion}
\label{S1:Conclude}

In this study, we measured the strength and direction of information flow between 28 level-1 SWS sector indices of the Chinese stock market, using transfer entropy. We constructed transfer entropy based information flow networks with the sector indices as the nodes and the net information flows between different sectors as the edges. In order to detect the structure of main information transfers among different sectors of the whole stock market, we adopted the maximum spanning arborescence technology to the information flow networks to extract important information transfer paths.

We considered first the outgoing and incoming MSAs of the information flow networks for the whole sample under investigation (see Fig.~\ref{Fig:MST:OI:whole}). We found that the \textit{composite} sector located at the center of the outgoing MSA as the information source to transmit information to other sectors. In contrast, the \textit{non-bank finance} sector was the root node of the incoming MSA, playing a role of the information sink in the stock market. By analyzing the MSAs in different years dynamically, we found that the \textit{non-bank finance}, \textit{bank}, \textit{non-ferrous metals}, \textit{textiles and apparel} sectors have appeared as high-degree root nodes. These sectors had more direct informative interactions with other sectors and played an important role in the information transfer networks of the Chinese stock market. Especially after the 2008 economic crisis, the two sectors (\textit{non-bank finance} and \textit{bank}) appeared as the highest-degree nodes in the outgoing MSAs, indicating that they played a crucial role during the economic recovery period.

We also investigated how the stock market turmoils affect the structure of the information flow networks. To this end, we analyzed the MSAs before, during and after the two main market turmoils around 2008 and 2015. We found that the degrees of the root sectors in all arborescences during market turmoils were higher than those in calm periods, except for the outgoing information flow MSA before the second turmoil, where the source node was the \textit{bank} sector. The somehow ``abnormal'' role played by the \textit{bank} sector was determined by the economic environment and trader behavior around 2015. For the maximal information flow paths of the pre-turmoil and post-turmoil MSAs, we found their total weights converged, which indicates that there is a similar information status of the market after a crisis.

We further investigated the specificity of source and sink sectors of the outgoing and incoming information flow MSAs by comparing their correlation coefficients with the SSEC returns. We found that the source sectors are less correlated with the SSEC and the sink sectors are more correlated with the SSEC, which means that information sources usually have idiosyncratic information and exhibit active traits, while information sinks usually exhibit passive traits in the market.

Overall, the structure of the information flow networks of the Chinese stock market changes along time. Information sources and sinks appear and also vary with time, showing the presence of the sector rotation phenomenon. We believe that these features observed in the Chinese stock market are very likely to hold in other stock markets, be they emerging or mature. 

Our work has several implications {for different market participants. More specifically, for policymakers such as the China Securities Regulatory Commission (CSRC) of the Chinese stock market, it will be useful to manage the market risk and control the contagion of risks. For example, during turbulent times, the CSRC should introduce policies to stabilize the root sector since during turbulent times, information sinks will be much riskier, while information sources may carry more contagious risk factors. For institutional investors, it provides clear direction in the process of asset allocation in different sectors according to the role of sectors in information flow networks. Furthermore, individual investors can also invest depending on the industry to which the stock belongs, for example, aggressive investors could buy stocks of the root sectors or sectors in maximal information flow paths which are the information source with high probability.} The work has also limitations. In particular, information sinks and sources may change during different time periods and differ in different markets. It is unclear if there are universal patterns in the appearance of information sinks and sources and the information flow paths. More likely, such patterns change with market circumstances and show adaptive behavior \cite{Lo-2004-JPM}.

\bibliographystyle{IEEEtran}


\newpage

\begin{IEEEbiography}[{\includegraphics[width=1in,height=1.25in,clip,keepaspectratio]{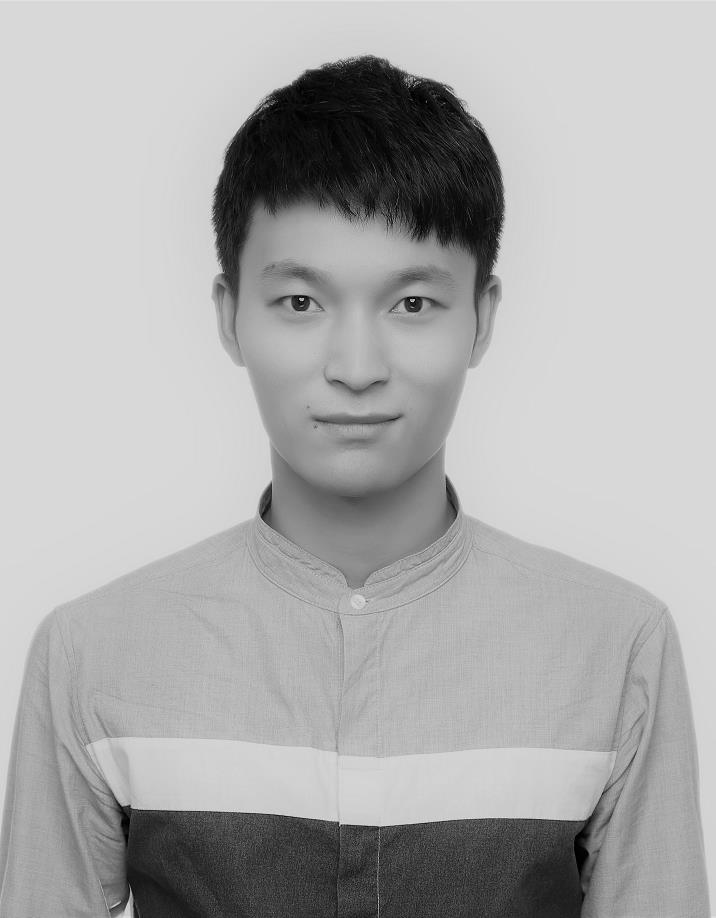}}]{Peng Yue} received his B.Sc. degree in mathematics and applied mathematics from East China University of Science and Technology. He is currently  pursuing the Ph.D. degree in management science and engineering at East China University of Science and Technology. He has been an Academic Guest in ETH Zurich from 2017 to 2019. His research interests include data science, statistics, finance and econophysics.
\end{IEEEbiography}

\begin{IEEEbiography}[{\includegraphics[width=1in,height=1.25in,clip,keepaspectratio]{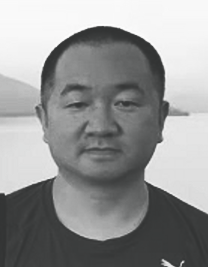}}]{Qing Cai} received his B.Sc. degree and M.Sc. degree in applied economics from East China University of Science and Technology, in 2002 and 2008, respectively. He is currently pursuing the Ph.D. degree with the Department of Finance at East China University of Science and Technology. His research interests include finance and econophysics.
\end{IEEEbiography}

\begin{IEEEbiography}[{\includegraphics[width=1in,height=1.25in,clip,keepaspectratio]{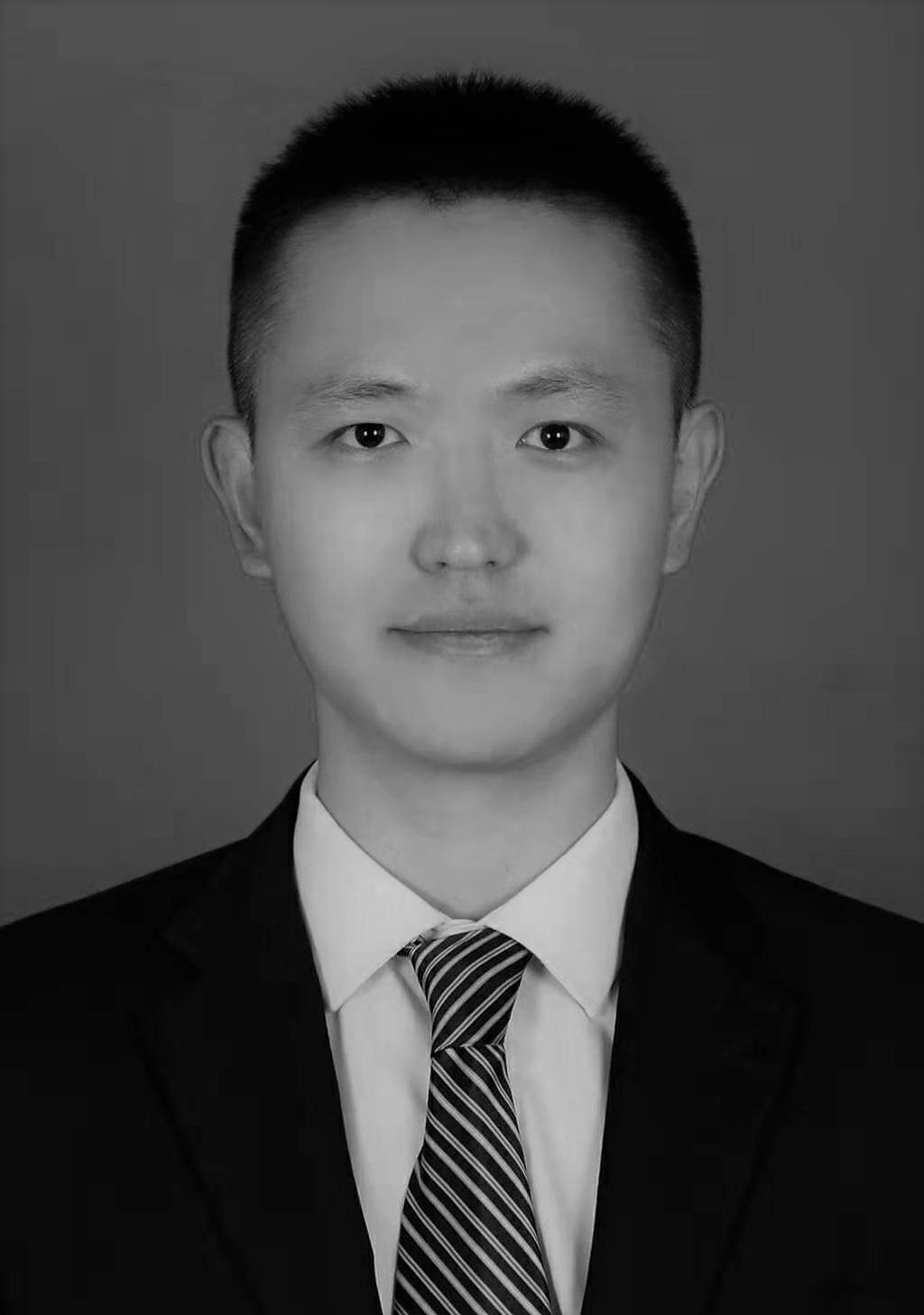}}]{Wanfeng Yan} received his B.Sc. degree from Peking University, M.Sc. degree and Ph.D. degree from ETH Zurich. He is currently the vice president of technology at Zhicang Technology (Beijing) and jointly affiliated to the Research Center for Econophysics at East China University of Science and Technology. His research interests include optimization, operations research, data science, statistics, finance and econophysics.
\end{IEEEbiography}

\begin{IEEEbiography}[{\includegraphics[width=1in,height=1.25in,clip,keepaspectratio]{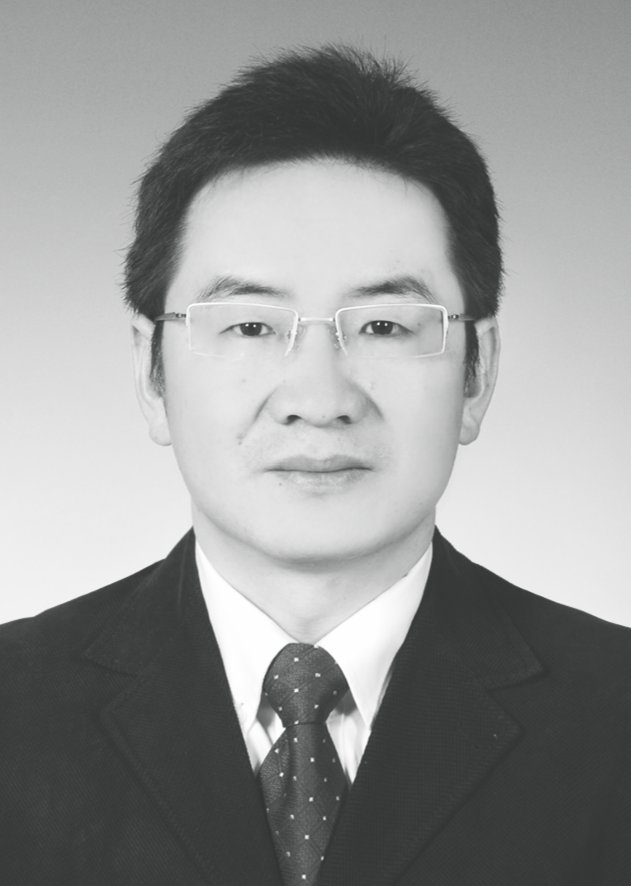}}]{Wei-Xing Zhou} received his Ph.D. in chemical engineering from East China University of Science and Technology in 2001. He did his post-doc at the University of California, Los Angeles in the group of Didier Sornette from 2001 to 2004. He joined East China University of Science and Technology in 2004 and was appointed as an Associated Research Professor in 2004 and a Professor of Finance, a Professor of Applied Mathematics, and a Professor of Management Science and Engineering in 2005. His research interests focus mainly on econophysics and sociophysics, especially multifractal analysis, financial bubbles and antibubbles, complex socioeconomic networks, and systemic financial risks.
\end{IEEEbiography}

\EOD
\end{document}